\documentclass{article}
\usepackage{amssymb,amsmath,amsfonts}
\usepackage{amsbsy}
\usepackage{epsfig}
\usepackage{enumerate}
\title{Transport of phase space densities through tetrahedral meshes using discrete flow mapping}
\author{Janis Bajars, David J.\ Chappell\vspace{1mm}\\
School of Science and Technology, Nottingham Trent University,\\ Clifton Campus, Clifton Lane,
        Nottingham, UK NG11 8NS\vspace{1mm}\\
Niels S{\o}ndergaard\vspace{1mm}\\
inuTech GmbH, F\"{u}rther Street, 90429 Nuremberg, Germany\vspace{1mm}\\
Gregor Tanner\vspace{1mm}\\
 School of Mathematical Sciences, University of Nottingham,\\ University Park, Nottingham, UK NG7 2RD}
\begin{document}
\maketitle
\begin{abstract}
Discrete flow mapping was recently introduced as an efficient ray
based method determining wave energy distributions in complex built
up structures. Wave energy densities are transported along ray
trajectories through polygonal mesh elements using a finite
dimensional approximation of a ray transfer operator. In this way
the method can be viewed as a smoothed ray tracing method defined
over meshed surfaces. Many applications require the resolution of
wave energy distributions in three-dimensional domains, such as in
room acoustics, underwater acoustics and for electromagnetic cavity
problems. In this work we extend discrete flow mapping to
three-dimensional domains by propagating wave energy densities
through tetrahedral meshes. The geometric simplicity of the
tetrahedral mesh elements is utilised to efficiently compute the ray
transfer operator using a mixture of analytic and spectrally
accurate numerical integration. The important issue of how to choose
a suitable basis approximation in phase space whilst maintaining a
reasonable computational cost is addressed via low order local
approximations on tetrahedral faces in the position coordinate and
high order orthogonal polynomial expansions in momentum space.
\end{abstract}

\section{Introduction}

Predicting the response of a complex vibro-acoustic system at mid-to
high frequencies is a long-standing challenge within the mechanical
engineering community \cite{Sestieri2013, Tanner2015p1}. Likewise,
characterizing the propagation of electromagnetic waves through
complex environments remains a formidable task, particularly with
respect to electromagnetic interference (EMI) and compatibility
(EMC) \cite{Gradoni2015, Montrose1998}. Asymptotic approximations
for high frequency waves lead to models based on geometrical optics,
where wave energy transport is governed by the underlying ray
dynamics and phase effects are neglected \cite{Runborg2007,
Chappell2013}. Directly tracking rays or swarms of trajectories in
phase space is often referred to as {\em ray tracing}, see for
example \cite{Cer01}. Methods related to ray tracing but tracking
the time-dynamics of beams or interfaces in phase space, such as
moment methods and level set methods, have been developed in Refs.~\cite{Osher2001}, \cite{Engquist2003} and \cite{Ying2006} amongst
others. They find applications in acoustics, seismology and computer
imaging, albeit restricted to problems with few reflections; for an
overview, see \cite{Runborg2007}.

Ray tracing and tracking methods can become inefficient in bounded
domains, or in general for problems including multiple scattering
trajectories and chaotic dynamics. Here, multiple reflections of the
rays and complicated folding patterns of the associated
level-surfaces often lead to an exponentially increasing number of
branches to be tracked. Instead of directly tracking trajectories,
we approach the problem here by tracking densities of rays as they
are transported along trajectories in phase space. Difficulties
owing to large numbers of reflections can thus be avoided
\cite{Chappell2013,Tanner2009}. More generally, high frequency wave
problems considered in this way become part of a wider class of
mass, particle or energy transport problems driven by an underlying
velocity field. Such problems arise in fluid dynamics \cite{CCMM04},
weather forecasting \cite{SR10} or in general in describing the
evolution of phase space densities by a dynamical system.

The transport of phase-space densities along a trajectory flow map
$\boldsymbol\varphi^{\tau}$ through time $\tau$ and space
$\mathbb{R}^d$ can be formulated in terms of a linear propagator
known as the Frobenius-Perron (FP) operator (see, for example,
\cite{Cvi12}). The action of this operator on a phase space density
$f$ may be expressed in the form
\begin{equation}\label{FPO}
\mathcal{L}^{\tau}f(\mathbf{X})=\int\delta(\mathbf{X}-\boldsymbol\varphi^{\tau}(\mathbf{Y}))f(\mathbf{Y})
\, \mathrm{d}\mathbf{Y},
\end{equation}
where $\mathbf{X}$ and $\mathbf{Y}$ are phase-space coordinates in
$\mathbb{R}^{2d}$. Solving such problems when $d>1$ and for
physically relevant systems is often considered computationally
intractable due to both the high dimensionality and the presence of
potentially complex geometries \cite{Lebot2002, Siltanen2007}. The
classical approach for dealing with such problems in applied
dynamical systems is to subdivide the phase space into distinct
cells and approximate the transition rates between these phase space
regions. A relatively simple approach whereby the phase space
densities in each of the cells are approximated by constants is
known as Ulam's method (see e.g.~\cite{Ding1996}). A detailed
discussion of the convergence properties of Ulam's method is given
in \cite{Bose2001} and \cite{Blank2002}. A number of related, but
more sophisticated, methods have been developed in recent years
including wavelet and spectral methods for the infinitesimal
FP-operator \cite{JK09, FJK11}, periodic orbit expansion techniques
\cite{Cvi12, Lippolis2010} and the so-called {\em Dynamical Energy
Analysis} (DEA) \cite{Tanner2009}. The modelling of many-particle
dynamics, such as protein folding, has been approached using short
trajectories of the full, high-dimensional molecular dynamics
simulation to construct reduced Markov models \cite{Noe09}. The
discrete ordinates method \cite{SC60, Thynell1998} is a related
approach with applications primarily in radiative heat transfer.
This method has been extended to multiple dimensions for relatively
simple geometries \cite{Thynell1998}.

In the following we focus on geometrical optics/acoustics models of
linear wave problems, although the methodology developed here can be
used in a more general context. Such models have been applied in
computer graphics since the mid-eighties \cite{Kayija1986} where the
rendering equation is used to transport the spectral radiance (of
light). The rendering equation has also been applied in room
acoustics \cite{Siltanen2007} leading to a method known as acoustic
radiance transfer. However, for its general application to complex
domains, simplifying assumptions are often necessary to obtain a
tractable numerical solution scheme. One commonly applied
simplification is the radiosity approximation, which leads to more
efficient computations since the density becomes independent of the
(phase space) direction coordinate. Similar techniques have been
applied in the realm of high-frequency structural vibrations
\cite{Lebot1998}.

Going a step further and assuming ergodicity and mixing of the
underlying ray dynamics, one can obtain a further simplified
modelling framework. Statistical Energy Analysis (SEA) (see for
example \cite{Lyon1969}, \cite{RL95} and \cite{Lafont2013}) is a
popular method of this kind in the structural dynamics community,
which is based on sub-dividing a structure into regions where the
above ray-dynamical assumptions are approximately valid. The result
is that the density in each subsystem is taken to be a single degree
of freedom in the model, leading to greatly simplified equations
based only on coupling constants between subsystems. A related
method developed in the electrical engineering community is the
random coupling model, which makes use of random field assumptions
(see \cite{Gradoni2014}). The disadvantage of these methods is that
the underlying assumptions are often hard to verify {\em a priori}
or are only justified when an additional averaging over `equivalent'
subsystems is considered. Possible generalisations and extensions of
SEA have been proposed in the works of Langley and Le Bot
\cite{Lebot2002, RL92, RL94, Lebot2006} amongst others.

In this paper we further develop the DEA methodology introduced in
\cite{Tanner2009}. Like the Ulam method, this approach is based on a
discrete representation of the FP operator. However, rather than
discretising the phase space volume, the FP operator is reformulated
as a phase-space boundary integral equation leading to an equivalent
model to the rendering equation with illumination points along the
entire boundary of the physical space. This boundary integral
equation is then written in a weak Galerkin form with a basis
approximation applied in both the position and momentum variable. In
Ref.~\cite{Tanner2009} the full domain is divided up into a number
of SEA-type subsystems and then the boundary integral formulation is
posed on this multi-domain system. In this way the level of
precision in the basis approximations gives rise to an interpolation
between SEA (at zeroth order) and full ray tracing (as the basis
order tends to infinity). Higher order basis approximations thus
relax the underlying ergodicity and quasi-equilibrium assumptions of
SEA. A more computationally efficient approach using a boundary
element method for the spatial approximation has been applied to
both two and three dimensional problems in \cite{Chappell2013} and
\cite{Chappell2012}, respectively. A major advantage of DEA is that
by removing the SEA requirements of diffusive wave fields
(equivalent to the ergodicity assumption) and quasi-equilibrium
conditions, the choice of subsystem division is no longer critical.

The modelling of three-dimensional problems using DEA was first
presented in \cite{Chappell2012}. However, the combination of high
dimensionality and costly quadrature routines including near
singularities meant that even performing the relatively low order
simulations presented in \cite{Chappell2012} required computation
times too long to give a viable numerical method. In order to
improve the efficiency of computing these multi-dimensional
integrals, the \textit{discrete flow mapping} (DFM) approach was
proposed in \cite{CTLS13, Wamot14}. DFM provides an efficient
numerical implementation of DEA on meshes and facilitates the
computation of phase space densities on complex two-dimensional
shell and plate-type structures by making use of the geometric
simplicity of typical mesh elements. The results in the
two-dimensional case have opened up the possibility of modelling of
large structures with millions of degrees of freedom \cite{ISMA14}
and point to the potential for much faster algorithms in three
dimensions. Preliminary work towards developing DFM for three
dimensional problems was discussed in a recent conference paper
\cite{ICSV15}. In this work we detail a DFM approach for tetrahedral
meshes and by extension, tetrahedralised three-dimensional
structures in general.

The paper is structured as follows. In Section 2 we outline the
governing operator equations for describing the evolution of a
phase-space density through a flow map between the faces of a
tetrahedral mesh. Section 3 details the discretisation of this
operator equation, and in particular, the efficient evaluation of
the discretised evolution operator using a combination of analytic
and spectral numerical integration methods. In Section 4 we present
numerical results for two examples; we consider one example with a
regular geometry and an exact solution for verification, but which
is typically unsuitable for a DFM model. The second example
considered was provided by an industrial collaborator and is a
tetrahedral mesh that was originally used for a finite element model
of a vehicle interior.

\section{Frobenius-Perron operator on a tetrahedral mesh}

\begin{figure}
\vspace{-1.5cm} \centering
\includegraphics[width=13cm]{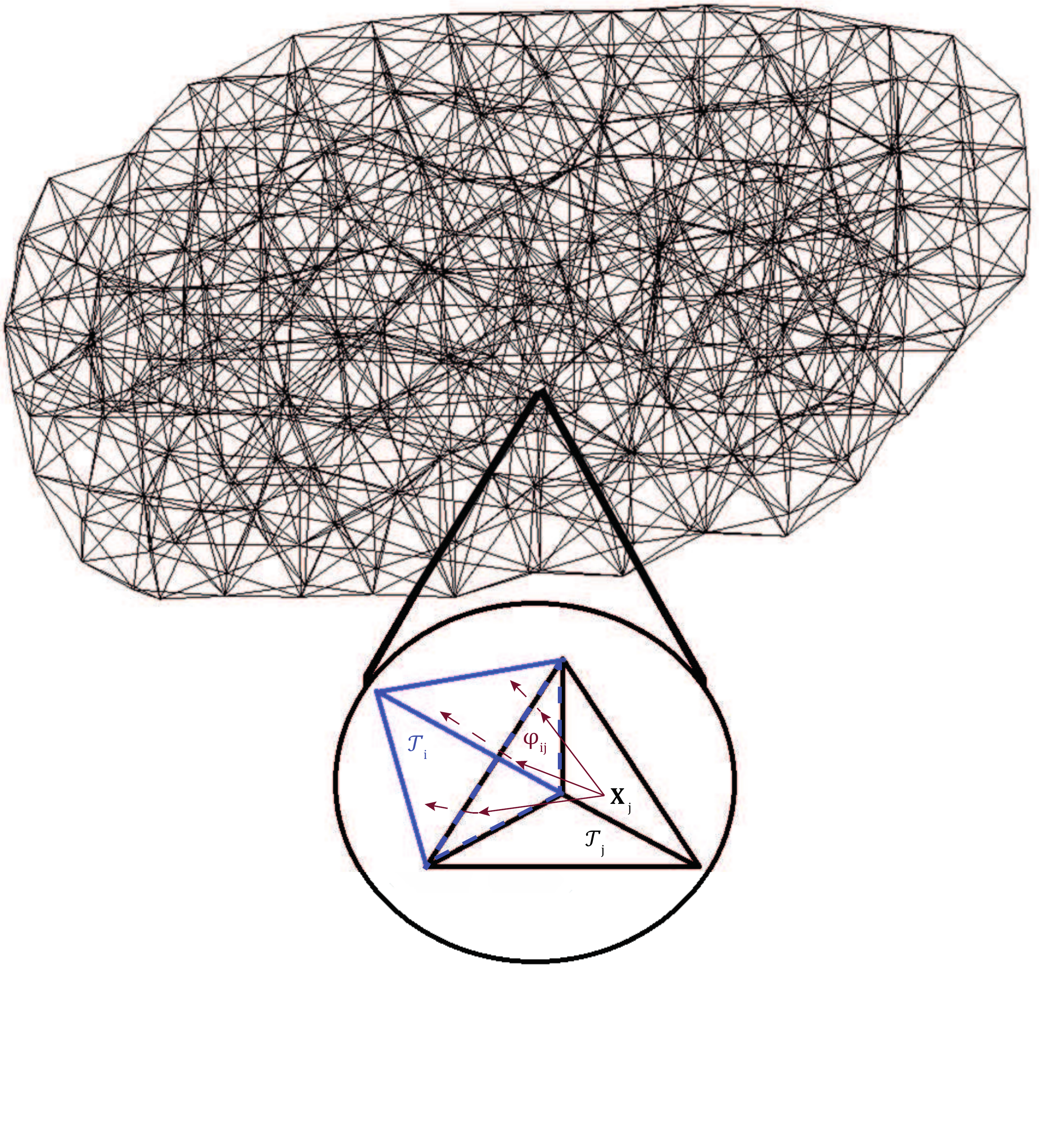}
\caption{\label{boundary-map}Discrete flow mapping on a mesh of a
vehicle interior showing the boundary flow map
$\boldsymbol\varphi_{ij}$ between the faces of a pair of adjacent
tetrahedra. (Online version in colour.)}
\end{figure}
Consider the propagation of a density  $f$ through a tetrahedral
mesh
$\mathcal{M}=\bigcup_{j=1}^{N}\mathcal{T}_{j}\subset\mathbb{R}^3$
consisting of $N$ tetrahedra $\mathcal{T}_{j}$, $j=1,...,N$, such as
depicted in Fig.~\ref{boundary-map}. Let us assume that the energy
of the trajectory flow is governed by piecewise constant
Hamiltonians of the form
$H_j(\mathbf{r},\mathbf{p})=c_j|\mathbf{p}|=1$ in $\mathcal{T}_j$,
where $c_j$ is the phase velocity for $\mathbf{r}\in\mathcal{T}_j$
and the momentum coordinate $\mathbf{p}$ lies on a sphere of radius
$c_j^{-1}$. This Hamiltonian is associated with the Helmholtz
equation with inhomogeneous wave velocity $c(\mathbf{r})$ (see
\cite{Runborg2007}). We restrict our discussion to scalar
propagation for simplicity; extensions to the vectorial wave
equations arising in elasticity and electromagnetics are possible.
In this case the wave propagation needs to be characterised by more
than one Hamiltonian per tetrahedron.

Denote the phase space on the boundary of the tetrahedron
$\mathcal{T}_j$ as $Q_j =
\partial\mathcal{T}_{j}\times D_{1/c_j}$, where the notation $D_{R}$ refers
to the open disk of radius $R$ and centre at the origin. Then the
associated coordinates are given by
$\mathbf{X}_{j}=[\mathbf{s}_j,\mathbf{p}_j] \in Q_j$ with
$\mathbf{s}_j\in\mathbb{R}^2$ parameterising
$\partial\mathcal{T}_j$, the boundary of the $j$th tetrahedron, and
$\mathbf{p}_j \in D_{1/c_j}$ parameterising the component of the
inward momentum (or slowness) vector tangential to
$\partial\mathcal{T}_j$. Next we define
$\boldsymbol\varphi_{ij}:Q_j\rightarrow Q_i$ to be the boundary flow
map, which takes a vector in $Q_j$ and maps it along the Hamiltonian
flow given by $H_j$ to a vector in $Q_i$, see Fig.~\ref{boundary-map}. Note that $\boldsymbol\varphi_{ij}$ is generally
only defined on a subset of $\partial\mathcal{T}_j$ and only maps to
a subset of $D_{1/c_i}$. The propagation of a density $f$ along the
boundary flow map $\boldsymbol\varphi_{ij}$ is therefore given by
the FP operator acting on this map as follows
\begin{align}\label{FPOtet2}
\mathcal{L}f(\mathbf{X}_{i})=\sum_j\int_{Q_j}\delta(\mathbf{X}_{i}-\boldsymbol\varphi_{ij}(\mathbf{X}_{j}))f(\mathbf{X}_{j}) \, \mathrm{d}\mathbf{X}_{j}.
\end{align}
The operator $\mathcal{L}$ describes propagation of $f$ along a
trajectory with endpoint on the boundary of tetrahedron
$\mathcal{T}_i$ and start point on the boundary of each neighbouring
or coincident tetrahedron $\mathcal{T}_j$. In cases where a
tetrahedral face is shared by the tetrahedra $\mathcal{T}_i$ and
$\mathcal{T}_j$ with $i\neq j$, then a reflection/transmission law
should be applied to specify the probability that the trajectory
endpoint lies in each of $\partial\mathcal{T}_i$ and
$\partial\mathcal{T}_j$. In order to include reflection/transmission
along with other physics such as dissipation (or mode conversion for
vectorial equations) we add a weighting function
$w_{i,j}(\mathbf{X}_{j})$, see Eq.~(\ref{FPOtetW}) below.

The stationary density $\rho(\mathbf{X}_{i})$ on $Q_i$, $i=1,...,N$,
due to an initial boundary distribution $\rho_0$ on $Q_j$,
$j=1,...,N$, is the density accumulated in the long time (many
reflection) limit. That is
\begin{equation}\label{NSum}
\rho(\mathbf{X}_{i})=\sum_{n=0}^\infty\mathcal{B}^{n}\rho_0(\mathbf{X}_{i}),
\end{equation}
where $\mathcal{B}^{n}$ describes the transport of the initial
density $\rho_0$ through $n$ reflections. Explicitly,
$\mathcal{B}^{n}$ is the $n$th iterate of the operator $\mathcal{B}$
given by
\begin{equation}\label{FPOtetW}
\mathcal{B}f(\mathbf{X}_{i})=\sum_j\int_{Q_j}w_{i,j}(\mathbf{X}_{j})\delta(\mathbf{X}_{i}-\boldsymbol\varphi_{ij}(\mathbf{X}_{j}))f(\mathbf{X}_{j})
\, \mathrm{d} \mathbf{X}_{j}.
\end{equation}

In this work we consider only the case of deterministic propagation
operators. Stochastic propagation may be described by replacing the
$\delta$-distribution in (\ref{FPOtetW}) with a finite-width kernel
(see for example Ref.~\cite{Chaos14}). If $w_{i,j}=1$, or more
generally, if there are no dissipative terms and the initial phase
space density is conserved, that is,
\[\sum_i\int_{Q_i} \left[\mathcal{B}^n\rho_0\right](\mathbf{X}_i) \, \mathrm{d}\mathbf{X}_i \, = \sum_i\int_{Q_i} \rho_0(\mathbf{X}_i) \, \mathrm{d}\mathbf{X}_i \, \quad \forall n, \]
then the operator $\mathcal{B}$ defined in (\ref{FPOtetW}) is of
Frobenius-Perron type as in (\ref{FPOtet2}) with a maximum
eigenvalue of 1. To obtain convergence of the sum in Eq.~(\ref{NSum}), a dissipative factor needs to be added; typically we
apply an absorption factor of the form $\exp(-\mu_j L)$, where $L$ is the length
of the trajectory and $\mu_j>0$ is the damping coefficient in
tetrahedron $\mathcal{T}_j$. We also consider an example where the
dissipative contribution is instead provided by an open boundary
region.

In the case where the sum in Eq.~(\ref{NSum}) converges, then
from the standard Neumann series result we have that the stationary
density $\rho(\mathbf{X}_{i})$ may be computed by solving the
following phase space boundary integral equation
\begin{equation}\label{GovIE}
(I-\mathcal{B})\rho(\mathbf{X}_{i})=\rho_0(\mathbf{X}_{i}).
\end{equation}
Note that since the trajectory flow only maps to neighbouring
tetrahedra, a matrix representation of $\mathcal{B}$ over the whole
of $\bigcup_{i=1}^{N}Q_{i}$ is in general sparse. In the next
section we design a discretisation strategy for efficiently
computing such a matrix representation of $\mathcal{B}$ and hence
numerically solving Eq.~(\ref{GovIE}).

\section{Discretisation}
In this section we describe a finite basis approximation of the
stationary density $\rho$ and the linear integral operator
$\mathcal{B}$ defined in (\ref{FPOtetW}). We provide an algorithm
for computing the entries of the resulting matrix representation $B$
of the operator $\mathcal{B}$. Furthermore, we describe a fast
semi-analytic and semi-spectral integration strategy for computing
the integrals arising in the definition of the entries in $B$.

\subsection{Finite basis approximation}\label{sec:basdisc}
We consider a finite dimensional approximation of the stationary
boundary density $\rho$ on $Q_j$ using a (product) basis expansion of the form
\begin{equation}\label{eq:FBApprox}
\rho(\mathbf{s}_j,\mathbf{p}_j) = \sum_{l=1}^{N_j} \sum_{n=0}^{N_p}
\sum_{m=-n}^{n} \rho_{(j,l,n,m)} b_{l}(\mathbf{s}_j)
\tilde{Z}_{n}^{m}(\mathbf{p}_j),
\end{equation}
where $N_j$ is the number of boundary elements on the tetrahedron
$\mathcal{T}_j$. For simplicity, we assume that each triangular face
forms a single element only and thus $N_j=4$ for all $j=1,\dots,N$.
We note however that the precision of the spatial approximation may
be improved by further sub-dividing the tetrahedral boundary faces.
Also, $N_p$ is the order of the basis expansion in the direction
coordinate $\mathbf{p}_j$. We apply orthonormal piecewise-constant basis functions $b_{l}$ in
the space coordinate $\mathbf{s}_j$, with support only on
$\bigtriangleup_{j,l}\subset\partial{\mathcal{T}_j}$, the
$l^{\text{th}}$ boundary element on $\mathcal{T}_j$. Hence
$$b_{l}(\mathbf{s}_j)=\frac{1}{\sqrt{|\bigtriangleup_{j,l}|}}$$
if $\mathbf{s}_j\in\bigtriangleup_{j,l}$ and is zero otherwise,
where we use $|\bigtriangleup_{j,l}|$ to denote the area of
$\bigtriangleup_{j,l}$. The functions $\tilde{Z}_{n}^{m}$ form an
orthonormal basis in the direction coordinate $\mathbf{p}_j \in
D_{1/c_j}$ and are given by
\begin{equation}\label{eq:MomBas}
\tilde{Z}_{n}^{m}(\mathbf{p}_j) = c_j
Z_{n}^{m}(\tilde{\mathbf{p}}_j),
\end{equation}
where $Z_{n}^{m}$ are the Zernike polynomials and
$\tilde{\mathbf{p}}_j:=(\varrho_j,\phi_j)\in[0,1)\times[0,2\pi)$ is
a re-scaling of $\mathbf{p}_j=(c_j^{-1}\varrho_j,\phi_j)$ to the
unit disk.

In fact, there are many possible candidates for an orthogonal basis
on a disc as outlined in Ref.~\cite{Boyd2011}. The main advantages
for the choice of Zernike polynomials here are their spectral
convergence for interpolating analytic functions and their relative
tractability in comparison with, for example, the Logan-Shepp ridge
polynomials \cite{Boyd2011}. In addition, only half as many degrees
of freedom are required to represent a complicated function on the
disk compared with a Chebyshev-–Fourier basis \cite{Boyd2011}. The
latter property results from the fact that the inner sum in
(\ref{eq:FBApprox}) only runs over the index values of the sum to
$N_p$, and that $Z_{n}^{m}\equiv0$ for $n-m$ odd.

The Zernike polynomials are defined as
\begin{align*}
Z_{n}^{m}(\tilde{\mathbf{p}}_j) &= R_{n}^{m}(\varrho_j) \cos(m\phi_j), \quad m \in \mathbb{Z}_{0}^{+},\\
Z_{n}^{m}(\tilde{\mathbf{p}}_j) &= R_{n}^{|m|}(\varrho_j)
\sin(|m|\phi_j), \quad m \in \mathbb{Z}^{-},
\end{align*}
where $n\in\mathbb{Z}_{0}^{+}$ and $R_{n}^{m}$ are polynomials of
the radial coordinate only. Note that $R_{n}^{m}\equiv0$ for $n-m$
odd, leading to the corresponding property for $Z_{n}^{m}$ described
above. The relative tractability of the Zernike polynomials stems
from the fact that their radial and angular dependence is separable
into terms that can be easily calculated. The radial part is defined
recursively via
\[
 R_{n}^{m}(\varrho) = \varrho \left( R_{n-1}^{|m-1|}(\varrho) + R_{n-1}^{m+1}(\varrho) \right) - R_{n-2}^{m}(\varrho),
 \quad R_{0}^{0}(\varrho) = 1, \quad R_{1}^{1}(\varrho) = \varrho,
\]
for any $\varrho\in[0,1]$ and the angular part is simply a
trigonometric function. For completeness, note that the
orthogonality property of Zernike polynomials (for $n-m$ even) is
given by
\[
 \int_{0}^{2\pi}\int_{0}^{1} Z_{n}^{m}(\tilde{\mathbf{p}}_j) Z_{n'}^{m'}(\tilde{\mathbf{p}}_j) \, \varrho_j\mathrm{d} \varrho_j\mathrm{d} \phi_j =
 \frac{\epsilon_{m'}\pi}{2n'+2}\delta_{n,n'}\delta_{m,m'},
\]
where $\epsilon_{m'}=1$ for all $m'\neq 0$ and $\epsilon_{0}=2$.

\subsection{Discretisation of the integral operator $\mathcal{B}$}
We apply a Galerkin projection of the operator $\mathcal{B}$
(\ref{FPOtetW}) onto the finite basis described above, which leads
to a matrix representation $B$ with entries
\begin{equation}\label{Bapprox}
B_{I,J}=\frac{2n'+2}{\epsilon_{m'}\pi} \int_{Q_j}
w_{i,j}(\mathbf{X}_{j})
\tilde{Z}_{n'}^{m'}(\boldsymbol\varphi_{p}(\mathbf{X}_{j}))
b_{l'}(\boldsymbol\varphi_{s}(\mathbf{X}_{j}))
\tilde{Z}_{n}^{m}(\mathbf{p}_j) b_{l}(\mathbf{s}_j) \, \mathrm{d}
\mathbf{X}_{j},
\end{equation}
where the subscripts $I$ and $J$ denote multi-indices
$I=(i,l',n',m')$ and $J=(j,l,n,m)$, respectively. The boundary map
$\boldsymbol\varphi_{i,j}=(\boldsymbol\varphi_s,\boldsymbol\varphi_p)$
is separated into spatial and directional components and the weight
function $w_{i,j}$ contains absorption and reflection/transmission
factors, as before. Once the entries of $B$ given in (\ref{Bapprox})
have been computed, the approximation of the stationary boundary
density $\rho$ is obtained by solving the linear system
corresponding to the discrete form of equation (\ref{GovIE}) for the
expansion coefficients $\rho_{(j,l,n,m)}$ in (\ref{eq:FBApprox}).

In order to write the entries of $B$ more explicitly in the form we
compute them, we substitute (\ref{eq:MomBas}) into (\ref{Bapprox})
and perform a change of variables $\varrho_j=\sin(\theta_j)$, where
$\theta_j\in[0,\pi/2)$ is the angle of the trajectory with respect
to the normal vector on $\bigtriangleup_{j,l}$. Furthermore, we
write out fully the factors in the weight $w_{i,j}$ and separate the
four integrals into two pairs to emphasise the relative simplicity
of the spatial integrand as follows:
\begin{align}\label{B4DInt}
\begin{split}
B_{I,J}
=\frac{2n'+2}{\epsilon_{m'}\pi\sqrt{|\bigtriangleup_{j,l}||\bigtriangleup_{i,l'}|}}
\int_{0}^{2\pi} \int_{0}^{\pi/2} \lambda_{i,j}(\theta_j,\phi_j)
Z_{n'}^{m'}(\tilde{\mathbf{p}}_i') Z_{n}^{m}(\tilde{\mathbf{p}}_j) \\
 \left[ \iint\limits_{\bigtriangledown_{j,l}} e^{-\mu_j L(\mathbf{s}_{j},\mathbf{s}'_{i})} \, \mathrm{d}\mathbf{s}_{j}\right]  \frac{\sin(2\theta_j)}{2} \, \mathrm{d}\theta_j
 \,\mathrm{d}\phi_j.
\end{split}
\end{align}
Here we have written
$\boldsymbol\varphi_{i,j}(\mathbf{X}_{j})=[\boldsymbol\varphi_{s}(\mathbf{X}_{j}),\boldsymbol\varphi_{p}(\mathbf{X}_{j})]=[\mathbf{s}_i',\mathbf{p}_i']$
for brevity and $L(\mathbf{s}_{j},\mathbf{s}'_{i})$ is the Euclidean
length of the trajectory connecting $\mathbf{s}_j$ and
$\mathbf{s}_i'$. Note that $\tilde{\mathbf{p}}_i'$ is simply
$\mathbf{p}_i'$ after re-scaling to the unit disc. Furthermore,
$\bigtriangledown_{j,l} \subseteq \bigtriangleup_{j,l}$ is the
triangular subset of points in $\bigtriangleup_{j,l}$ that are
mapped to $\bigtriangleup_{i,l'}$ by
$\boldsymbol\varphi_{ij}(\mathbf{X}_{j})$. Note that this subset
depends on the direction of propagation $(\theta_j,\phi_j)$. The
restriction of the spatial integration to $\bigtriangledown_{j,l}$
results from the basis function $b_{l'}$ in (\ref{Bapprox}) setting
the integrand to zero for trajectories where
$\mathbf{s}_i'\notin\bigtriangleup_{i,{l'}}$. In addition, we have
separated the weight function into the product of the
reflection/transmission probability $\lambda_{i,j}(\theta_j,\phi_j)$
and the absorption factor $\exp(-\mu_j L)$.

In the case of flat-faced polyhedra, the Euclidean length
$L(\mathbf{s}_{j},\mathbf{s}'_{i})$ is a linear function of
$\mathbf{s}_j$ and hence the spatial integral over
$\bigtriangledown_{j,l}$ can be computed analytically. We are then
left with an integral over the unit disc to compute numerically. The
strategy for performing the exact integration over the tetrahedral
mesh element faces provides information about the regularity of the
remaining integrand as a function of the direction coordinates. We
describe this strategy in the next section and use it to motivate
the design of a spectrally converging quadrature scheme for the
outer integrals.

\subsection{Computation of the matrix entries $B_{I,J}$}
\begin{figure}
\begin{center}
\includegraphics[trim=1cm 0cm 1cm 0cm,clip=true,width=0.45\textwidth]{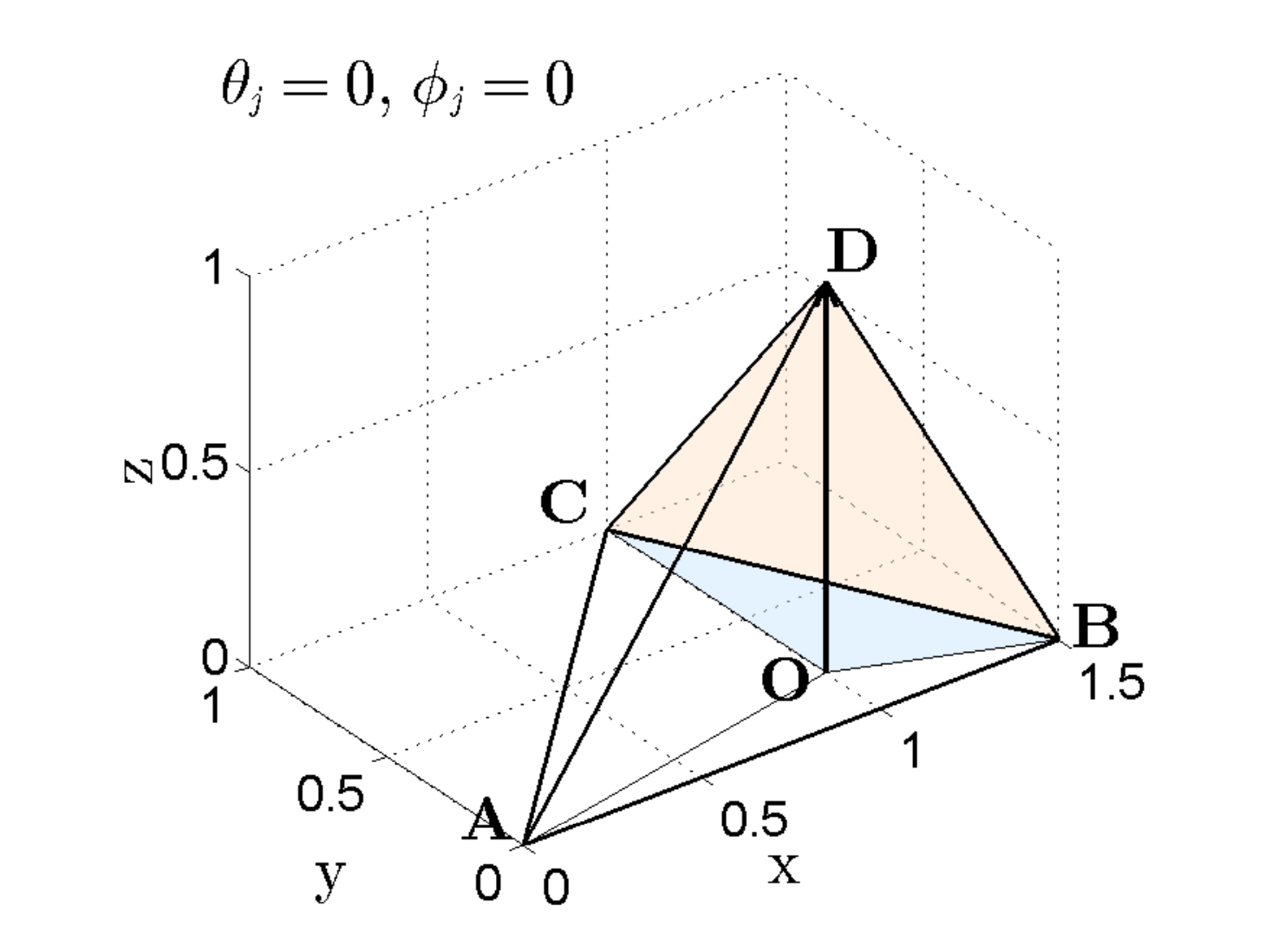}
\includegraphics[trim=1cm 0cm 1cm 0cm,clip=true,width=0.45\textwidth]{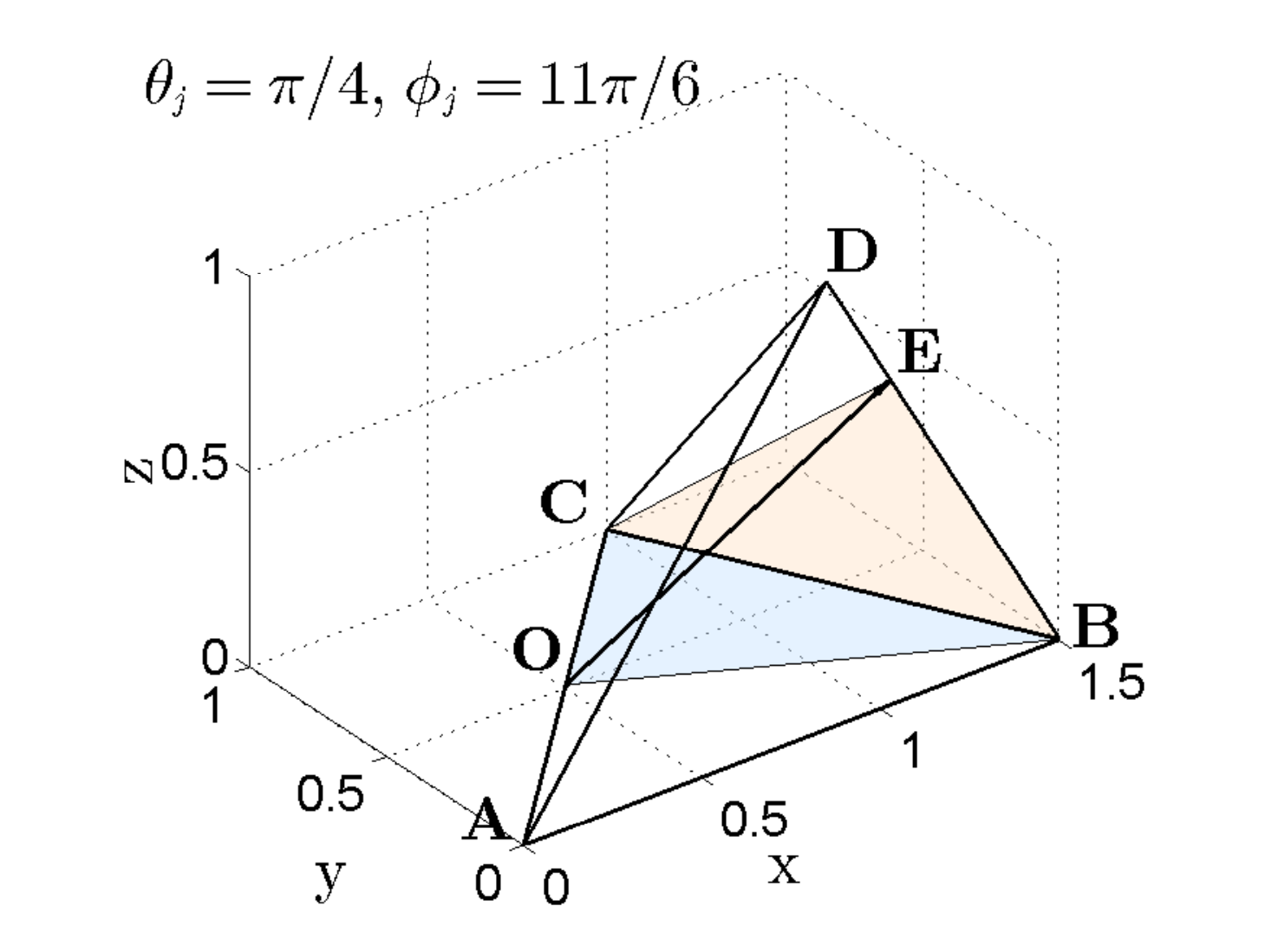} \\
\includegraphics[trim=1cm 0cm 1cm 0cm,clip=true,width=0.45\textwidth]{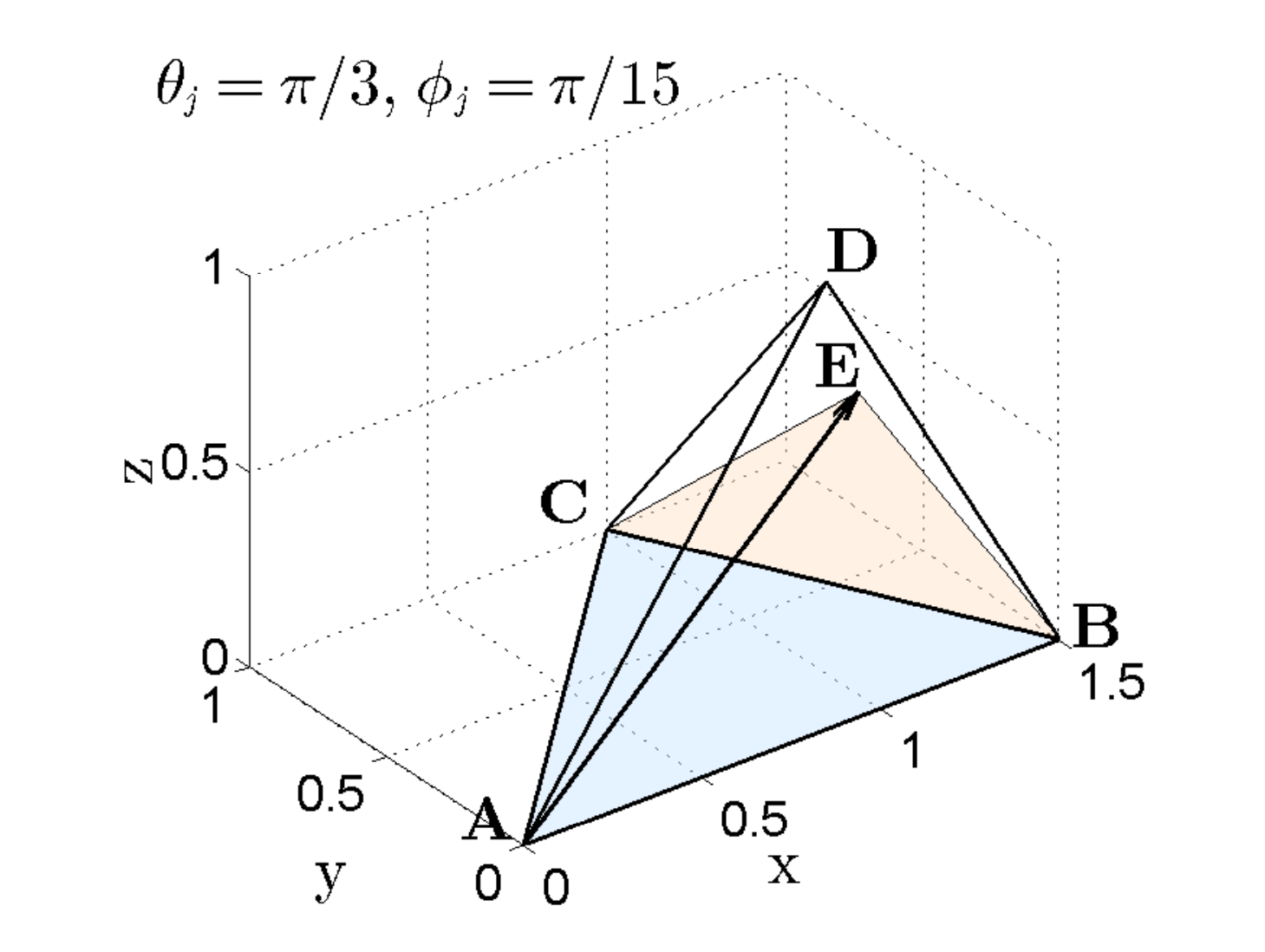}
\includegraphics[trim=1cm 0cm 1cm 0cm,clip=true,width=0.45\textwidth]{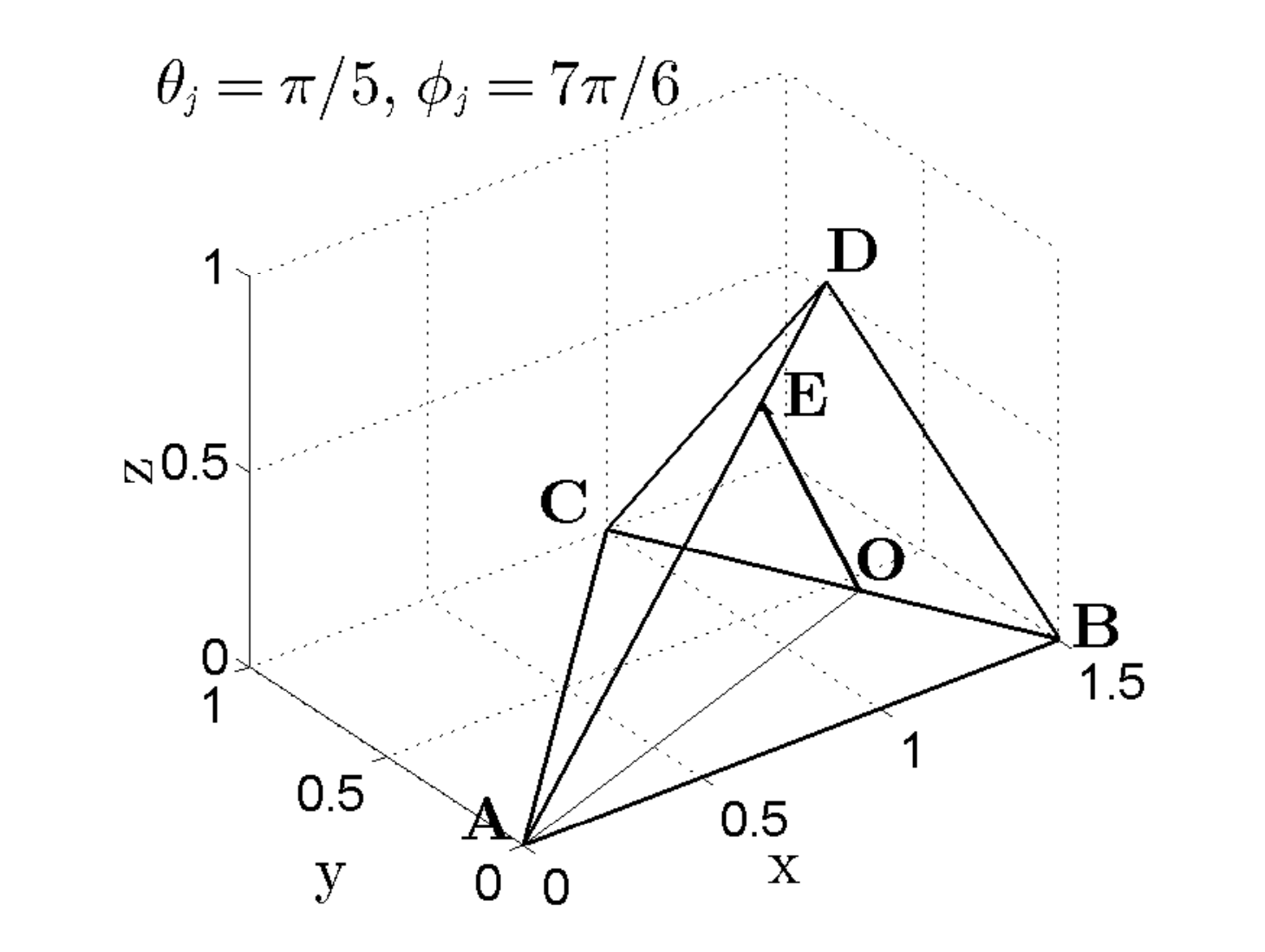}
\end{center}
\caption{Tetrahedron with vertices $\mathbf{A}$, $\mathbf{B}$,
$\mathbf{C}$ and $\mathbf{D}$. The shaded sub-regions show the
admissible sets of start and end points for rays starting from face
$ABC$ and arriving in face $CBD$. The set of admissible start points
within triangle $ABC$ is equal to the spatial integration domain
$\bigtriangledown_{j,l}$. Each of the four sub-plots shows these
regions for a different fixed outgoing ray direction
$(\theta_j,\phi_j)$. The trajectory of maximum length $L_{max}$ is
indicated by the black arrow. Upper left: the set of admissible
trajectory end-points is equal to the whole of $CBD$. Upper right:
the sets of admissible start and end points are subsets of $ABC$ and
$CBD$, respectively. Lower left: the set of admissible trajectory
start points is equal to the whole of $ABC$. Lower right: the sets
of admissible start and end points are both empty. (Online version
in colour.)}\label{TetraPlots}
\end{figure}

In this section we describe the necessary steps for computing the
four-dimensional integrals given in (\ref{B4DInt}). For the purposes
of illustration, consider a single tetrahedron with vertices
$\mathbf{A}$, $\mathbf{B}$, $\mathbf{C}$ and $\mathbf{D}$ as shown
in Fig.~\ref{TetraPlots}. The figure shows the admissible sets of
start and end points for rays travelling in a specified fixed
direction from face $ABC$ to face $CBD$. Each tetrahedral face has a
prescribed local coordinate system with origin at one of the
vertices (vertex $A$ in Fig.~\ref{TetraPlots}). The local $x$-axis
is taken to be aligned with one of the edges (edge $AB$ in
Fig.~\ref{TetraPlots}) and the local $z$-axis is taken parallel to
the inward normal vector of the face. We rotate and translate each
tetrahedron in the mesh, together with its neighbouring tetrahedra,
such that the local face coordinates used for the spatial
integration variables coincide with the global Cartesian coordinates
as shown in Fig.~\ref{TetraPlots}.

\begin{figure}
\begin{center}
\includegraphics[width=0.65\textwidth]{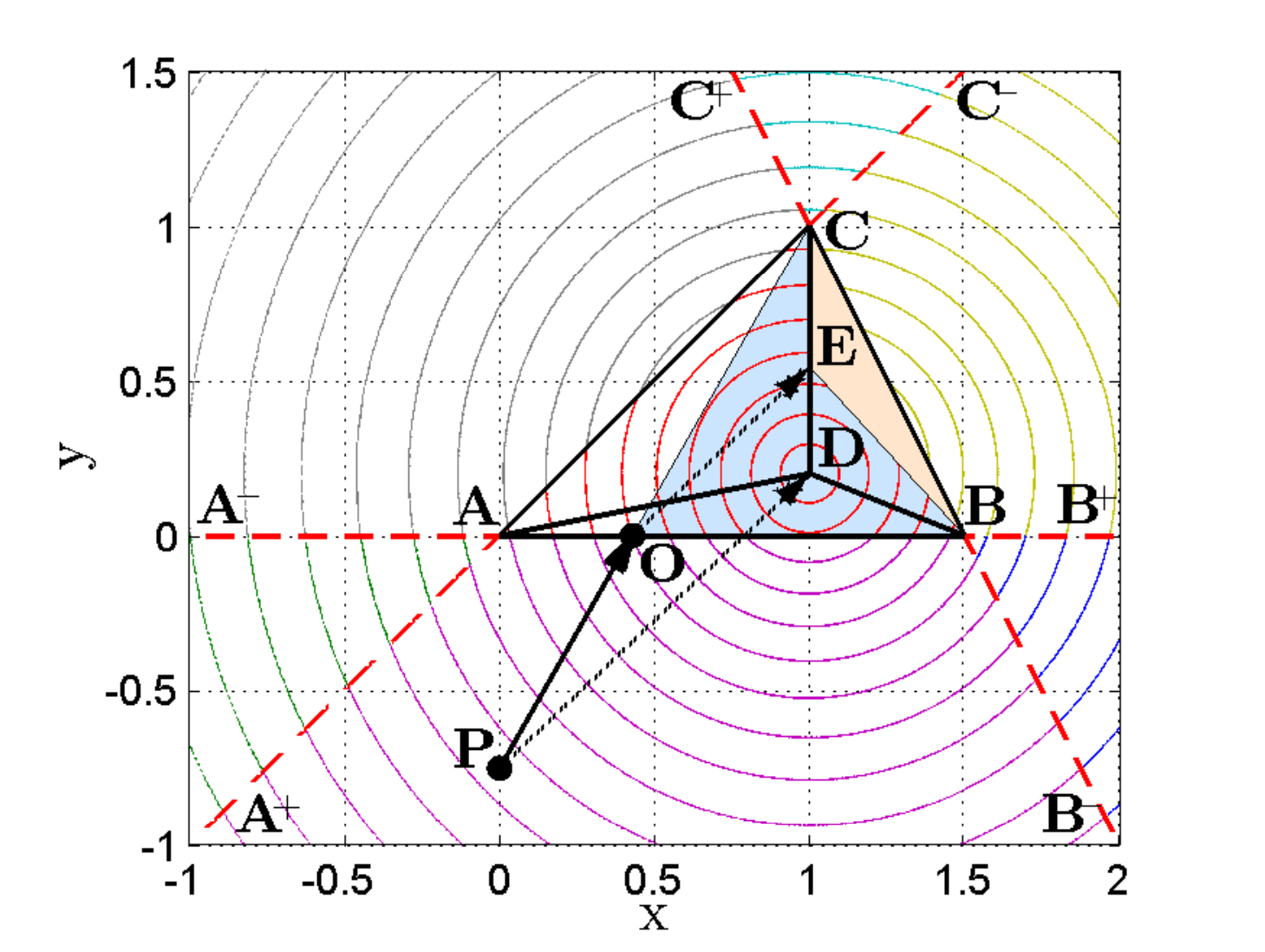}
\end{center}
\caption{Tetrahedron with vertices $\mathbf{A}$, $\mathbf{B}$,
$\mathbf{C}$ and
$\mathbf{D}=({D}_1,{D}_2,{D}_3)$ viewed from
above. The dashed lines ${A^{-}B^{+}}$, ${B^{-}C^{+}}$ and
${C^{-}A^{+}}$ extend the three edges of triangle $ABC$ and
demonstrate the subdivision of the $xy$-plane into seven
sub-regions. For given direction $(\theta_j,\phi_j)$, the point
$\mathbf{P}$ connects the $xy$-plane with vertex $\mathbf{D}$. If
$\mathbf{P}$ lies outside the triangle $ABC$, then the point
$\mathbf{O}$ is obtained by projecting $\mathbf{P}$ onto the
boundary of $ABC$. Such points $\mathbf{P}$ span the $xy$-plane via
circles of radius $D_3\tan(\theta_j)$ and origin $(D_1,D_2)$.
(Online version in colour.)}\label{xyCircles}
\end{figure}

We first find the triangular region $\bigtriangledown_{j,l}\subseteq
\bigtriangleup_{j,l}$ over which the spatial integration takes
place, for a given direction $(\theta_j,\phi_j)$. Consider the upper
left plot of Fig.~\ref{TetraPlots}, where the outgoing direction is
parallel to the normal vector, that is, $\theta_j=0$. Here the point
$\mathbf{O}$ forms the triangle $\bigtriangledown_{j,l}=OBC \subset
ABC$ over which to perform the spatial integration. Outside this
region ($OBC$) rays parallel to the normal direction $\mathbf{OD}$
will not approach face $BCD$, but instead approach one of the other
two faces, $ABD$ or $CAD$. However, the point $\mathbf{O}$ may not
be located inside the triangle $ABC$, but rather on the boundary of
$ABC$ as shown in the other three plots of Fig.~\ref{TetraPlots}.
The region $\bigtriangledown_{j,l}$ is therefore determined by
finding the location of the point $\mathbf{O}$. In order to do this,
first consider the point $\mathbf{P}$ in the local $xy$-plane such
that the vector $\mathbf{PD}$ coincides with the direction vector
specified by $(\theta_j,\phi_j)$, as illustrated in
Fig.~\ref{xyCircles}; this figure shows the tetrahedron from
Fig.~\ref{TetraPlots} viewed from above. We see that each direction
value $(\theta_j,\phi_j)$ specifies a unique point
$\mathbf{P}\in\mathbb{R}^2$. For example, when the angle $\theta_j$
is fixed but the angle $\phi_j$ varies from zero to $2\pi$, the
points $\mathbf{P}$ lie on circles of radius $D_3\tan(\theta_j)$ and
have origin $(D_1,D_2)$, where $(D_1,D_2,D_3)$ are the local
coordinates of the point $\mathbf{D}$. We extend the three edges of
$ABC$ as illustrated by the dashed lines in Fig.~\ref{xyCircles} and
indicate the end-point limits of these infinitely extended dashed
lines by $\mathbf{A}^{\pm}$, $\mathbf{B}^{\pm}$ and
$\mathbf{C}^{\pm}$. The lines ${A^{-}B^{+}}$, ${B^{-}C^{+}}$ and
${C^{-}A^{+}}$ divide $\mathbb{R}^2$ into seven regions, and the
region in which the point $\mathbf{P}$ lies prescribes one of seven
possible locations for the point $\mathbf{O}$; the point
$\mathbf{O}$ can be located either inside the triangle $ABC$, or on
one of its three vertices, or along one of its three sides (not
including the vertices).

In the simplest case, the point $\mathbf{P}$ already lies on or
within the triangle $ABC$ and then $\mathbf{O}=\mathbf{P}$. On the
other hand, if $\mathbf{P}$ lies in one of the three triangular
areas, ${A^{-}AA^{+}}$, ${B^{-}BB^{+}}$ or ${C^{-}CC^{+}}$, then the
point $\mathbf{O}$ is obtained by projecting $\mathbf{P}$ onto the
vertex shared with triangle $ABC$. If the point $\mathbf{P}$ lies in
one of the three quadrilateral areas, e.g.~${A^{+}AB B^{-}}$, then
the point $\mathbf{O}$ is obtained by projecting the point
$\mathbf{P}$ onto the edge  shared with triangle $ABC$. This
projection is taken along the line connecting the point $\mathbf{P}$
to the vertex of triangle $ABC$ not on the shared edge. In the
illustrative example shown in Fig.~\ref{xyCircles}, the point
$\mathbf{O}$ is defined by the intersection of the lines $AB$ and
$PC$.

In addition to finding the domain $\bigtriangledown_{j,l}$ for the
spatial integrals in (\ref{B4DInt}), we also need to find the
maximal ray length between the two faces under consideration for a
prescribed fixed direction $(\theta_j,\phi_j)$. In other words, we
find
$$L_{max}=\max\left\{L(\mathbf{s}_{j},\mathbf{s}'_{i})\right\},$$ where the distance
function $L$ is expressed in the local coordinate system of
$\bigtriangledown_{j,l}$.  For example, consider the tetrahedron
$ABCD$ in the upper-left plot of Fig.~\ref{TetraPlots}. Here,
$L_{max}=D_3$ and we place the origin of the local coordinate system
of triangle $OBC=\bigtriangledown_{j,l}$ at the vertex $\mathbf{B}$
such that the $x$-axis coincides with edge $BC$. More generally,
when the point $\mathbf{O}$ lies inside the triangle $ABC$, then the
ray of maximum length $L_{max}$ (indicated by a black arrow in
Fig.~\ref{TetraPlots}) will intersect the vertex $\mathbf{D}$. If
instead the point $\mathbf{O}$ lies on an edge of $ABC$ that is not
on the destination face $CBD$, then the ray of maximum length will
approach one of the edges of the destination face not on $ABC$ as
depicted in the upper right plot of Fig.~\ref{TetraPlots}. When the
point $\mathbf{O}$ coincides with the vertex $A$ of $ABC$ (not on
the destination face $CBD$) as shown in the lower left plot of
Fig.~\ref{TetraPlots}, then the longest ray will approach a point
$\mathbf{E}$ inside the destination face and the spatial integration
domain $\bigtriangledown_{j,l}= \bigtriangleup_{j,l}$. Finally, if
the point $\mathbf{O}$ lies on an edge or vertex of the receiving
face $CBD$, the region $\bigtriangledown_{j,l}=\emptyset$ and there
is no need to compute $L_{max}$ since the corresponding contribution
to the matrix entry $B_{I,J}$ is zero. Note therefore that the
direction coordinate space is also divided into seven distinct
sub-regions according to the seven possible locations for the point
$\mathbf{O}$ on triangle $ABC$. For three of these sub-regions, the
corresponding spatial integral will be zero as in the latter case
described above.

Once we have obtained the integration domain
$\bigtriangledown_{j,l}$ and the maximal ray length $L_{max}$, we
can proceed with the analytical evaluation of the spatial integrals
appearing in (\ref{B4DInt}). These integrals have a relatively
simple form since the distance function
$L(\mathbf{s}_{j},\mathbf{s}'_{i})$ is a linear function of the
local coordinates of triangle $\bigtriangledown_{j,l}$. Considering
again the tetrahedron $ABCD$ in the upper-left plot of
Fig.~\ref{TetraPlots} and taking $h$ to be the minimum distance from
the point $\mathbf{O}$ to the edge $BC$, then in this case the
spatial double-integral is given by
\[
\iint\limits_{OBC} e^{-\mu_j L(\mathbf{s}_{j},\mathbf{s}'_{i})} \,
\mathrm{d}\mathbf{s}_{j} = \frac{h|\mathbf{BC}|}{(\mu_j L_{max})^2}
\left( \mu_j L_{max} - 1 + e^{-\mu_j L_{max}}\right), \quad \mu_j >
0.
\]
In the case $\mu_j=0$, the integrand simplifies to unity and the
integral is simply the area of the domain $\bigtriangledown_{j,l}$.

In order to complete the evaluation of the matrix entries $B_{I,J}$
in (\ref{B4DInt}), we need to find the reflected/transmitted ray
directions $\tilde{\mathbf{p}}_i'=(\sin(\theta'_{i}),\phi'_{i})$ and
compute the reflection/transmission probabilities
$\lambda_{i,j}(\theta_j,\phi_j)$, before (numerically) integrating
over $(\theta_j,\phi_j)$. We apply Snell's Law to relate the
reflection angle  $\theta_r\in[0,\pi/2)$  to the transmission angle
$\theta_t\in[0,\pi/2)$ via
\[
\sin(\theta_t) = \frac{c_i}{c_j} \sin(\theta_r).
\]
The reflection angle $\theta_r$ corresponds to a specular reflection
of the incoming ray. If $i=j$ then $\theta'_i=\theta_r$, otherwise
we have $\theta'_i=\theta_t$. Once we have found $\theta'_i$, then
the azimuthal angle $\phi'_{i}$ must be represented in the local
coordinates of each face of the tetrahedral mesh using linear
transformations of the incoming ray direction. We find that the
reflection/transmission probability function is dependent only on
reflection/transmission angles $\theta_{r}$ and $\theta_{t}$. For
acoustic waves the transmission probability is given by
\[
\lambda_t(\theta_j,\phi_j) =
\frac{4z_iz_j\cos(\theta_r)\cos(\theta_t)}{(z_i\cos(\theta_r) +
z_j\cos(\theta_t))^2},
\]
where the specific acoustic impedance $z_j=\rho^{f}_jc_j$ is the
product of the fluid density $\rho_j^{f}$ and the propagation speed
$c_j$ in tetrahedron $\mathcal{T}_j$. Then we have
$\lambda_{i,j}=\lambda_t$ if $i\neq{j}$ and
$\lambda_{i,i}=1-\lambda_t$.

Consider a quadrature rule for the integration over the direction
coordinates $(\theta_j,\phi_j)$. For each pair $(\theta_j,\phi_j)$
specified by the quadrature rule, we carry out the following five
steps:
\begin{enumerate}
\item Determine the triangular region $\bigtriangledown_{j,l}$.
\item Compute the maximal length of rays travelling from face $\bigtriangleup_{j,l}$ to face $\bigtriangleup_{i,{l'}}$.
\item Compute the spatial (double) integral over the triangular area $\bigtriangledown_{j,l}$ analytically.
\item Find the reflection/transmission angles $(\theta'_{i},\phi'_{i})$ in the local face coordinates of $\mathcal{T}_i$.
\item Compute the Zernike polynomials $Z_{n}^{m}(\tilde{\mathbf{p}}_j)$ and $Z_{n'}^{m'}(\tilde{\mathbf{p}}_i')$, and the reflection/transmission function $\lambda_{i,j}(\theta_j,\phi_j)$.
\end{enumerate}
The numerical integration strategy is further detailed in the next
section.

\subsection{Subdivision and parametrisation for the spectral quadrature scheme}\label{sec:IntSubReg}\begin{figure}
\begin{center}
\includegraphics[width=0.65\textwidth]{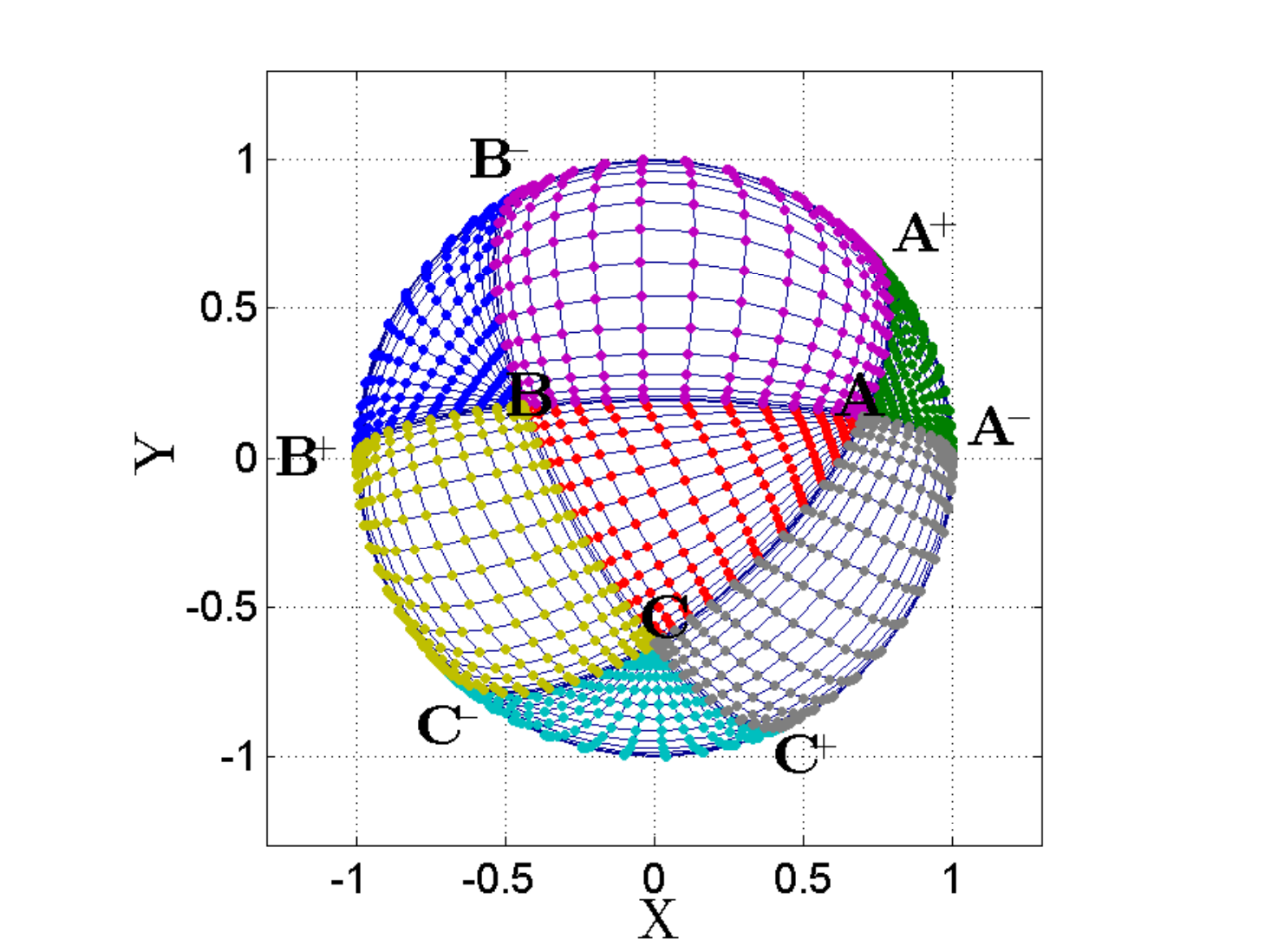}
\end{center}
\caption{Seven subregions of the direction coordinate over the unit
disc associated to the tetrahedron $ABCD$ presented in Figures
\ref{TetraPlots} and \ref{xyCircles}. (Online version in
colour.)}\label{SubRegions}
\end{figure}

In this section we describe a subdivision and parametrisation of the
direction space $(\theta_j,\phi_j)\in[0,\pi/2)\times[0,2\pi)$ in
order to obtain a spectrally convergent quadrature scheme for the
integrals with respect to $(\theta_j,\phi_j)$ appearing in
Eq.~(\ref{B4DInt}). As described in the previous section, the seven
regions in the local $xy$-plane indicated in Fig.~\ref{xyCircles}
are associated with seven regions in the direction space on the unit
disc as shown in Fig.~\ref{SubRegions}, where the labelled points
correspond to the points with the same labels shown in
Fig.~\ref{xyCircles}. Note that the lines ${A^{-}B^{+}}$,
${B^{-}C^{+}}$ and ${C^{-}A^{+}}$, which divide the seven regions,
are great circles of the unit sphere depicted in
Fig.~\ref{SubRegions}. The spatial integral (as the function of
direction) is smooth inside each of these subregions, but is only
continuous along the lines dividing the subregions. This property
serves as a good motivation for subdividing the integration in the
direction coordinate into seven subregions as indicated in
Fig.~\ref{SubRegions}.

We now describe a suitable mapping and parameterisation for any of
the seven subregions of the direction space illustrated in
Fig.~\ref{SubRegions}. The parametrisation of any triangular or
rectangular region on the upper hemisphere can be achieved using a
map of the form
\begin{equation}\label{Parametr}
\mathbf{f} (\xi,\eta) = \mathbf{a} + \mathbf{b} \xi + \mathbf{c}
\eta + \mathbf{d} \xi \eta \in \mathbb{R}^3, \quad
(\xi,\eta)\in[-1,1]^2,
\end{equation}
with
\begin{align*}
\mathbf{a} &= \frac{1}{4} (\mathbf{v}_1 + \mathbf{v}_2 + \mathbf{v}_3 + \mathbf{v}_4),\\
\mathbf{b} &=\frac{1}{4} (\mathbf{v}_2 - \mathbf{v}_1 + \mathbf{v}_3 - \mathbf{v}_4), \\
\mathbf{c} &= \frac{1}{4} (\mathbf{v}_3 + \mathbf{v}_4 - \mathbf{v}_1 - \mathbf{v}_2), \\
\mathbf{d} &= \frac{1}{4} (\mathbf{v}_3 - \mathbf{v}_4 +
\mathbf{v}_1 - \mathbf{v}_2),
\end{align*}
and where $\mathbf{v}_1$, $\mathbf{v}_2$, $\mathbf{v}_3$ and
$\mathbf{v}_4$ are four vertices on the upper-hemisphere. For a
triangular region with vertices $\mathbf{v}_1$, $\mathbf{v}_2$ and
$\mathbf{v}_3$, we simply set $\mathbf{v}_4=\mathbf{v}_1$. Then we
map $\mathbf{f}(\xi,\eta)$ onto the upper-hemisphere using
\cite{Boal2008}
\[
\mathbf{g}(\xi,\eta) =
\frac{\mathbf{f}(\xi,\eta)}{|\mathbf{f}(\xi,\eta)|}.
\]
Writing the entries of vectors $\mathbf{f}=(f_1,f_2,f_3)^T$ and
$\mathbf{g}=(g_1,g_2,g_3)^T$, then we compute the direction
$(\theta_j,\phi_j)$ as functions of $(\xi,\eta)$ via the standard
relations for spherical coordinates:
\begin{align}\label{sphpol}
\begin{split}
 \theta_j(\xi,\eta) & = \arccos(g_3(\xi,\eta)), \\
 \phi_j(\xi,\eta) & = \mbox{mod} \left( \arctan \left( \frac{f_2(\xi,\eta)}{f_1(\xi,\eta)} \right),2\pi\right).
 \end{split}
\end{align}
In addition, the Jacobian of this transformation including the
$\sin(2\theta_j)/2$ factor appearing in (\ref{B4DInt}) is given by
\[
{J}(\xi,\eta) = \frac{f_3}{|\mathbf{f}|^4}\:
\mathrm{det}\left[\mathbf{f}\hspace{2mm}\frac{\partial\mathbf{f}}{\partial\xi}\hspace{2mm}\frac{\partial\mathbf{f}}{\partial\eta}
\right].
\]

Thus all sub-regions of direction space are reduced to two
dimensional integrals over the square $[-1,1]^2$ and, in principle,
any quadrature method can be applied. In our computations we
consider a 2D adaptive Clenshaw-Curtis method. Clenshaw-Curtis
quadrature converges spectrally if the integrand is sufficiently
smooth. The splitting into seven sub-regions described above is
however, not always sufficient to ensure this smoothness. If there
is a change in propagation speed so that $c_i\neq c_j$ then the
integration variable should be changed to the transmission direction
in order to preserve the smoothness \cite{HighOrd2D}. Another reason
that a loss of regularity of the integrand may occur is that the
spherical coordinate relation (\ref{sphpol}) leads to singularities
in the $\phi_j$-dependent terms of the derivatives of the integrand
in (\ref{B4DInt}) when $\theta_j=0$. However, for sufficiently
regular tetrahedral meshes, we observe spectral convergence of the
Clenshaw-Curtis quadrature despite the singular behaviour near
$\theta_j=0$. This is due to the fact that the Zernike polynomials
$Z_{n}^{m}(\varrho_j,\phi_j)$ are equal to zero at $\theta_j=0$ for
any $m\neq0$ (recall $\varrho_j=\sin(\theta_j)$). Furthermore,
$Z_{n}^{0}(0,\phi_j)=(-1)^{n/2}$ and is thus independent of
$\phi_j$. For irregular tetrahedral meshes containing long slender
tetrahedra, the singularity may still cause numerical issues. In
this case, a change of variables using functions that are flat at
$\theta_j=0$ (that is, their derivatives vanish at $\theta_j=0$) can
be introduced to remove the singularity in higher derivatives of
integrand \cite{DR84}.

\section{Numerical results}
In this section we present results for two numerical examples.
Firstly, we consider one-dimensional trajectory propagation in a
cuboid, since here we can compare our result with both an exact
geometric solution and an averaged wave solution. Secondly, we
simulate the high frequency response of a vehicle cavity to a point
source excitation. Through both numerical results we demonstrate the
efficiency of DFM on tetrahedral meshes and the convergent behaviour
of the solution as the Zernike polynomial direction basis order is
increased.


\subsection{Verification for a cuboid cavity}

As a first example we consider a density distribution inside a
cuboid $(x,y,z)\in(0,\ell)\times(-0.5,0.5)\times(-0.5,0.5)$. We
prescribe a constant ray density along the boundary surface of the
cuboid at $x=0$ with a fixed inward direction taken along the normal
direction $(1,0,0)$. At all other boundaries of the cuboid we prescribe a homogeneous Neumann (or sound hard) boundary condition. This leads to a one-dimensional solution along the $x$-axis, which is independent of $y$ and $z$. Due to the
geometric simplicity, this example possesses both an exact
geometrical optics solution and an exact wave equation solution that
we will compare against the numerical DFM solutions for different
orders of Zernike polynomial basis approximation.

The exact solution $u$ for the associated wave problem is given by
solving a two-point Neumann boundary value problem for the Helmholtz
equation with complex wavenumber $k=\omega/c + i \mu/2$. Here
$\omega$ is the angular frequency, $c$ is the wave speed and
$\mu=\eta \omega /(2c)$ is a frequency-dependent dissipation rate
with (hysteretic) loss factor $\eta$. It is reasonably
straightforward to show that the solution with a unit Neumann
condition at $x=0$ and a homogeneous (Neumann) condition at $x=\ell$
is given by a sum of left and right travelling plane waves as
\begin{equation}\label{eq:WSol}
 u(x;\omega) = \frac{1}{2k\sin(k\ell)}(e^{ik(x-\ell)}+e^{ik(\ell-x)}).
\end{equation}
Assuming that $u$ defines a velocity potential in a fluid of density
$\rho^f$, then the acoustic energy density is given by
$\rho^f\omega^2|u|^2/c^2$ and the averaged acoustic energy density
over a frequency band $[\omega-\Delta
\omega/2,\omega+\Delta\omega/2]$ can be calculated via
\begin{equation}\label{eq:AvgWSol}
\rho_{\omega}(x) = \frac{\rho^f}{c^2\Delta \omega \,
}\int_{\omega-\Delta\omega/2}^{\omega+\Delta\omega/2}
q^2|u(x;q)|^2 \, \mathrm{d} q.
\end{equation}

The corresponding ray tracing model arises by transporting a source
density $\tilde{\rho}_{0}$ from $x=0$ towards $x=\ell$, where it is
reflected back towards $x=0$. After being transported though $n$
reflections at both $x=\ell$ and $x=0$ (after returning), the ray
density travelling from left to right is given by
\[
 \rho_+^n(x) = e^{-\mu(2\ell n + x)} \tilde{\rho}_{0}, \quad n=0,1,2\dots
\]
Likewise, the ray density travelling from right to left at the point
$x$ after $n$ reflections at $x=\ell$ is given by
\[
 \rho_-^n(x) = e^{-\mu(2\ell n - x)} \tilde{\rho}_{0}, \quad n=1,2\dots
\]
The final stationary density $\rho^{*}(x)$ is accumulated from the
contributions from both directions after each reflection and leads
to a geometric series solution of the form
\begin{align}\label{eq:ExactSol}
\rho^{*}(x) = \sum_{n=0}^{\infty} \rho_+^{n}(x)+ \sum_{n=1}^{\infty}
\rho_-^{n}(x)  = \frac{e^{-\mu x}+e^{-\mu (2\ell-x)}}{1-e^{-2\mu
\ell}}\tilde{\rho}_0.
\end{align}
A source density $\tilde{\rho}_{0}$ corresponding to the boundary
condition for the averaged wave solution (\ref{eq:AvgWSol}) can be
found by setting $\rho^*(0)=\rho_{\omega}(0)$. Applying this
condition and combining (\ref{eq:AvgWSol}) and (\ref{eq:ExactSol})
at $x=0$ leads to
$$\tilde{\rho}_{0}=\tanh(\mu\ell) \rho_{\omega}(0).$$
Note that for large $\omega$ and hence large $\mu$,
$\tanh(\mu\ell)\sim1$ and we have simply that $\tilde{\rho}_{0}\sim
\rho_{\omega}(0)$. Physically, this corresponds to the solution
being dominated by the initial plane wave travelling from left to
right, since the high damping leads to a negligible contribution
from the returning reflected waves. In this high frequency limit,
one can also derive a frequency independent source density as
\[
\rho_{\omega}(0)\sim  \frac{\rho^f}{1+\eta^2/16}.
\]

In the 3D DFM simulations we prescribe a constant initial density
value $\tilde{\rho}_0$ across the face of the cuboid at $x=0$ with
fixed direction parallel to the surface normal, that is we lift
$\tilde{\rho}_0$ to phase space via
\begin{equation}\label{eq:RhoNeumann}
\rho_0 = \tilde{\rho}_0 \frac{\delta(\varrho_j)}{2 \pi \varrho_j}.
\end{equation}
Recall $\varrho_j=\sin(\theta_j)$ and note that the delta
distribution has been normalised for polar coordinates so that
$$\int_0^{2\pi}\int_0^1\frac{\delta(\varrho_j)}{2\pi \varrho_j}\varrho_j \, \mathrm{d}\varrho_j \, \mathrm{d}\phi_j=1.$$
The initial density (\ref{eq:RhoNeumann}) is then projected onto the
finite basis approximation (\ref{eq:FBApprox}) in order to perform
the numerical simulations.

\begin{figure}
\begin{center}
\includegraphics[trim=0cm 0cm 0cm 0cm,clip=true,width=0.9\textwidth]{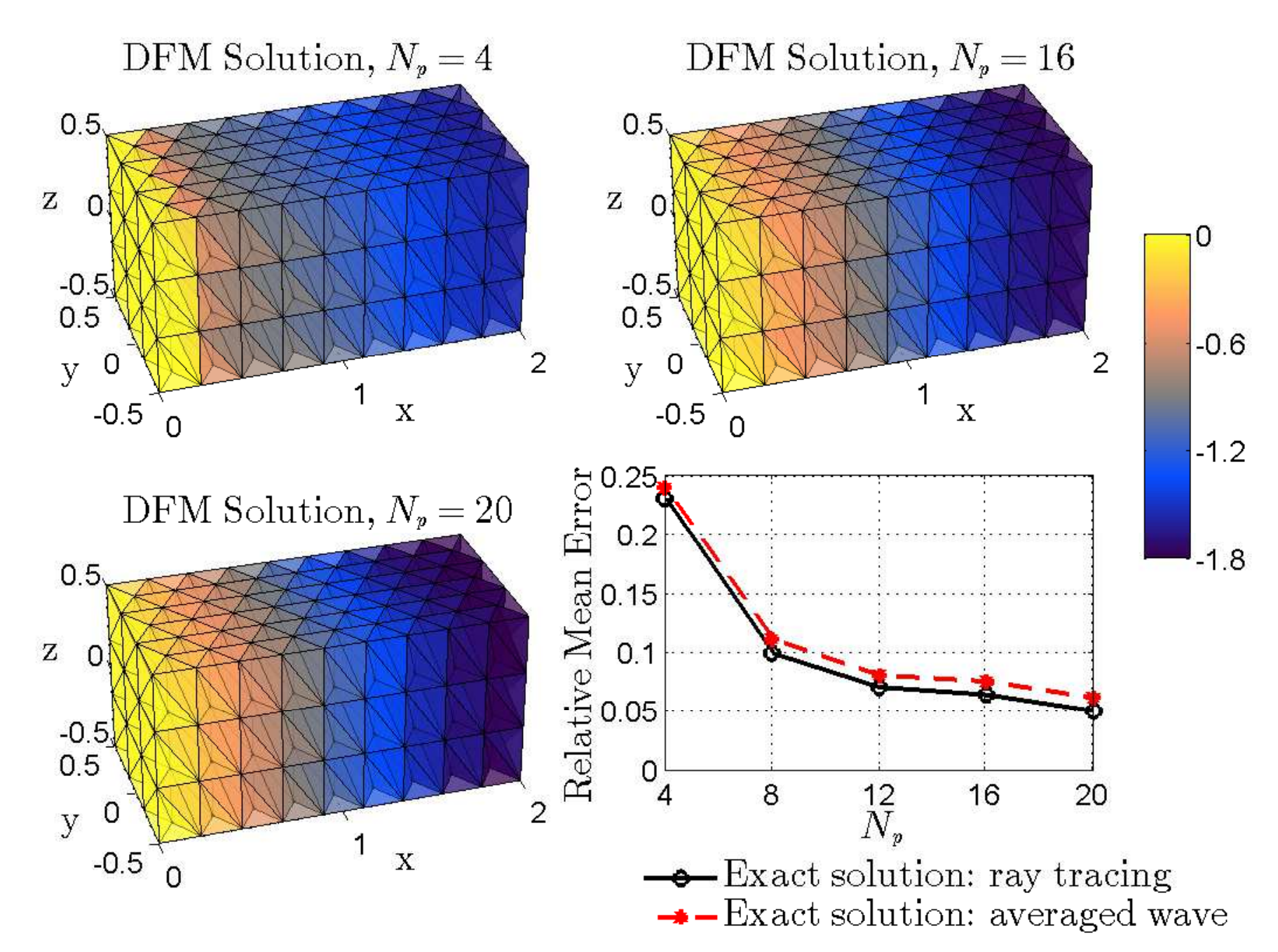}
\end{center}
\caption{The interior density $\rho_{\mathcal{T}}$ (see Eq.~(\ref{eq:IntDens})) in a cuboid mesh with parameter values $N=972$,
$\omega=200\pi$, $\eta=0.004$, $\rho_j^{f}=1$ and $c_j=1$ for all
$j=1,\ldots,N$. The first three plots show the DFM simulation
results for $\log(\rho_{\mathcal{T}}(\mathbf{r}))$ with $\mathbf{r}$ chosen as the centroid of each tetrahedron and with Zernike polynomial direction basis order $N_p=4$ (upper left),
$N_p=16$ (upper right) and $N_p=20$ (lower left). The lower right plot shows the relative mean error of
DFM simulations for different approximation orders where the exact
solution is given by either the exact ray density solution
(\ref{eq:ExactSol}) or the averaged exact wave solution
(\ref{eq:AvgWSol}). (Online version in colour.)}\label{fig:Cuboid}
\end{figure}

The numerical results for a cuboid with $\ell=2$ are shown in
Fig.~\ref{fig:Cuboid}. We employ a tetrahedral mesh with $N=972$
elements and choose the fluid density and the propagation speed to
be $\rho^{f}_j=c_j=1$ for all $j=1,\ldots,N$. We also take
$\omega=200\pi$ and apply a loss factor of $\eta=0.004$. The
numerical integrals are evaluated using a spectrally convergent
Clenshaw-Curtis rule on appropriately defined sub-regions. Figure
\ref{fig:Cuboid} shows the interior stationary density
$\rho_{\mathcal{T}}$ plotted on a logarithmic scale. This interior
density may be computed at $\mathbf{r}\in\mathcal{T}_j$ by
projecting the stationary boundary density $\rho$ onto the interior
position space via
\begin{equation}\label{eq:IntDens}
 \rho_{\mathcal{T}}(\mathbf{r}) = \frac{1}{c_j^3}\sum_{l=1}^{4} \int_{\bigtriangleup_{j,l}} \rho(\mathbf{s}_j,\mathbf{p}_j) \frac{\cos(\vartheta(\mathbf{s}_j,\mathbf{r}))}{|\mathbf{r}-\mathbf{r}_{s_j}|^2} e^{-\mu_j |\mathbf{r}-\mathbf{r}_{s_j}|} \, \mathrm{d}
 \mathbf{s}_j.
\end{equation}
Here $\mathbf{r}_{s_j}\in\mathbb{R}^3$ are the Cartesian coordinates
of the point $\mathbf{s}_j\in\partial\mathcal{T}_j$ and
$\vartheta(\mathbf{s}_j,\mathbf{r})\in[0,\pi/2)$ is the angle
between the normal vector to $\bigtriangleup_{j,l}$ (pointing into
$\mathcal{T}_j$) and the direction vector
$\mathbf{r}-\mathbf{r}_{s_j}$.

The first three plots of Fig.~\ref{fig:Cuboid} show the interior
density results computed at the centroid of each tetrahedron. The
DFM result with $N_p=4$ shown in the upper-left plot clearly differs
from the two higher order computations shown in the upper-right and
lower-left plots. The discrepancy is most clear towards the right of
the cavity where the energy density has not decayed to the level
shown in the other plots. The plots with $N_p=16$ and $N_p=20$ are
very similar suggesting that the solution is reasonably well
converged on this particular mesh. The lower-right plot shows the
relative mean error, that is, the mean absolute error divided by the
mean of the exact solution taken over all tetrahedra in the mesh.
The errors are plotted for various approximation orders from $N_p=4$
to $N_p=20$ and suggest convergence towards a solution with around
$5\%$ error compared to the exact ray density solution
(\ref{eq:ExactSol}) and around $6\%$ error with reference to the
averaged exact wave solution (\ref{eq:AvgWSol}). The exact solutions
themselves are not plotted in Fig.~\ref{fig:Cuboid}, since they have
a very similar appearance to the higher order DFM results. The
averaged exact wave solution was calculated using a frequency band
of $\Delta\omega=20\pi$ and the larger error in this case can be
attributed to the geometrical optics approximation of the wave
problem. The fact that the errors in both cases are fairly similar
suggests that a ray-based model is appropriate for this problem.

The results presented could be improved by using a finer tetrahedral
mesh and increasing the Zernike polynomial approximation order until
the error appears to be saturating as above. Alternatively, one
could consider subdividing the tetrahedral faces and applying the
piecewise constant spatial basis over smaller boundary elements of
the tetrahedral boundary as discussed in Sect.~\ref{sec:basdisc}, or
employing higher order basis approximations in space as discussed
for two-dimensional problems in Ref.~\cite{HighOrd2D}. We also note
that a relatively low damping level with loss factor $\eta=0.004$
has been applied and that for larger damping values the solution
would decay more quickly from left to right. In this case a finer
spatial approximation using one of the methods described above would
be required to achieve the accuracy level shown in
Fig.~\ref{fig:Cuboid}. Using an error tolerance of $10^{-4}$ for the
numerically evaluated integrals, the computational times for the
results presented above are around 90s for the $N_p=4$ result, 100
minutes for $N_p=16$ and 300 minutes for $N_p=20$.

The cuboid example considered in this section is highly directional
with propagation only along the $x$--direction and thus poses a
great challenge for modelling with DFM, which approximates the whole
direction space using a smooth basis. In fact, DFM is particularly
suited to modelling complex structures with many reflections where
the dependence of the solution on the direction of propagation will
be far smoother. Such an example will be considered in the next
section. Despite its limitations for such highly directive problems,
we note that DFM was able to reproduce the qualitative solution
behaviour to a reasonable level of accuracy.

\subsection{Application to a tetrahedral mesh of a vehicle cavity}

\begin{figure}
\begin{center}
\includegraphics[trim=0cm 0cm 0cm 0cm,clip=true,width=1\textwidth]{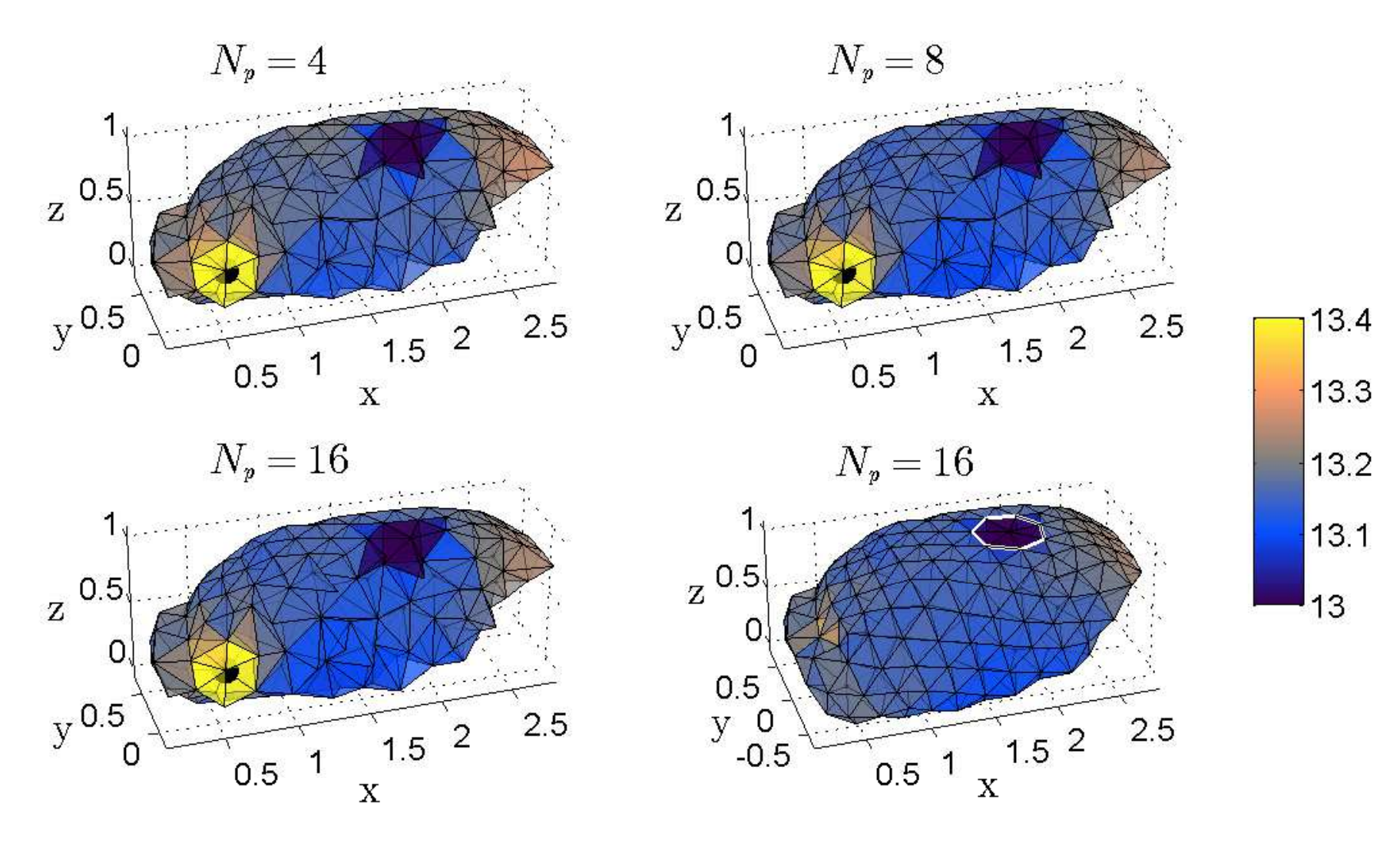}
\end{center}
\caption{The DFM results for the interior density $\rho_{\mathcal{T}}$ in a vehicle
cavity mesh with parameter values $N=1300$, $\omega=200\pi$, $\rho^{f}_j=1$, $c_j=1$ and $\mu_j=0$ for all $j=1,\ldots,N$. The plotted quantity is $\log(\rho_{\mathcal{T}}(\mathbf{r}))$ with $\mathbf{r}$ chosen as the centroid of each tetrahedron. The boundary of an opening in the
upper part of the cavity is indicated in white on the lower right
plot. The other three plots all show the cavity sliced approximately
through the centre; only tetrahedra with centroids in the half-space
$y>0$ are plotted. The source point at
$r_0=(0.56467,0.024773,0.28073)$ is therefore visible in these three
plots and is indicated by a black spot. The results are shown with
Zernike polynomial direction basis order $N_p=4$ (upper left),
$N_p=8$ (upper right) and $N_p=16$ (both lower plots). (Online
version in colour.)}\label{fig:Caravan}
\end{figure}

In this section we present the results of DFM simulations in a
vehicle cavity discretised by $N=1300$ tetrahedral mesh elements as
shown in Fig.~\ref{fig:Caravan}. The cavity is excited with an
interior point source and is assumed to have a specularly reflecting (outer) boundary where energy is conserved, except for a small open region in the roof through which energy is lost. We place the source point $\mathbf{r}_0$ at
one of the mesh vertices $\mathbf{r}_0=(0.56467,0.024773,0.28073)$,
which is located towards the front of the vehicle cavity. The point
source velocity potential gives rise to an acoustic energy density
on the boundary of any neighboring tetrahedra $\mathcal{T}_j$ of the
form
\begin{equation}\label{eq:SourceRho}
\rho_0(\mathbf{s}_j,\mathbf{p}_j;\mathbf{r}_0) =
\frac{\rho^f_j\omega^2 \cos(\vartheta(\mathbf{s}_j,\mathbf{r}_0))
\delta(\mathbf{p}_j-\mathbf{p}_0(\mathbf{s}_j,\mathbf{r}_0))}{16\pi^2c_j|\mathbf{r}_0-\mathbf{r}_{s_j}|^2}
e^{-\mu_j |\mathbf{r}_0-\mathbf{r}_{s_j}| },
\end{equation}
where $\mathbf{r}_{s_j}\in\mathbb{R}^3$ are the Cartesian
coordinates of $\mathbf{s}_j\in\partial\mathcal{T}_j$ as before and
$\mathbf{p}_0(\mathbf{s}_j,\mathbf{r}_0)\in D_{1/c_j}$ is the tangential part
of the momentum vector with direction
$\mathbf{r}_0-\mathbf{r}_{s_j}$. In this example we set $\mu_j=0$
for all $j=1,\ldots,N$; note that the dissipation required for the
sum in Eq.~(\ref{NSum}) to converge is provided via the losses
through the open boundary region instead. The boundary of the
opening is illustrated in white in the lower right plot of
Fig.~\ref{fig:Caravan}.

In order to perform the DFM computation, we project the source
boundary density (\ref{eq:SourceRho}) onto the finite basis
approximation (\ref{eq:FBApprox}). We fix $\rho^{f}_j=c_j=1$, and
compute all numerical integrals to single precision using spectrally
convergent Clenshaw-Curtis quadrature. We consider Zernike
polynomial basis orders $N_p=4,8$ and $16$. A cross-section of the
cavity with the source point indicated by a black spot is shown in
the first three plots of Fig.~\ref{fig:Caravan}, while the final
plot (lower-right) illustrates the whole cavity.  In the first three
plots the cavity is sliced approximately through the centre in the
$y$-direction, and only tetrahedra whose centroids are located in
the half-space $y>0$ are plotted. The interior density is computed
at the centroid of each tetrahedron and plotted on a logarithmic
scale as before. The computational results for approximation orders
$N_p=8$ and $N_p=16$ are very similar and illustrate the convergence
of the DFM result on this particular mesh. These higher order
computations indicate a stronger shadowing effect underneath the
open boundary region, compared with the result for $N_p=4$.
Increasing the order of the approximation in direction space
therefore better captures the loss of energy through the
non-reflecting region, causing less energy to propagate into the
cavity immediately underneath to the opening.

We note that the 3D DFM simulations presented here are vastly more
computationally efficient than the 3D DEA simulations presented in
\cite{Chappell2012}. The largest computation presented in this
section ($N_p=16$) took approximately $10$ hours to run using a
MATLAB implementation of the code. A similar computation on a
simpler cavity using only a second order basis approximation in
momentum took several weeks using the 3D DEA method reported in
\cite{Chappell2012}. It should be noted that DFM is also easily
parallelisable using a simple parameter sweep over the tetrahedra
$j=1,\ldots,N$ and hence the computational time could be improved
further by performing parallelised computations within a high
performance programming language. We also note that although the
problems modelled in this work have used conforming tetrahedral
meshes, an extension to non-conforming meshes would simply be a case
of ensuring that the trajectory flows are associated to the
appropriate destination tetrahedra. Furthermore, the feasibility of
the exact spatial integration method relies only on linearity and
not on the tetrahedral geometry itself; extensions to general convex
polyhedra would be possible but more complicated to implement.

\section{Conclusions}

We have extended the DFM approach described in \cite{CTLS13} to
model wave energy densities in three-dimensional domains. In DFM,
the densities to be computed are transported along ray trajectories
through tetrahedral mesh elements using a finite dimensional
approximation of a ray transfer operator. The relative geometric
simplicity of the tetrahedral mesh elements has been exploited to
efficiently compute the entries in the matrix representation of the
discretised ray transfer operator. In particular, each matrix entry
requires the evaluation of a four-dimensional integral; two
integrals have been evaluated analytically, and the other two have
been computed using spectrally convergent quadrature methods.
Numerical results have been presented to verify the methodology, and
to demonstrate its convergence and efficiency in practice for a
full-scale vehicle cavity mesh provided by an industrial
collaborator.

\section*{Acknowledgments}
Support from the EU (FP7-PEOPLE-2013-IAPP grant no.~612237 (MHiVec))
is gratefully acknowledged. The authors also thank Dr Gang Xie from
CDH AG for providing the vehicle cavity mesh data.

\bibliographystyle{ieeetr}

\end{document}